\documentclass[graybox]{svmult}
\usepackage{ifthen}
\newboolean{articlemode}
\setboolean{articlemode}{true}
\newboolean{latexmode}
\setboolean{latexmode}{true}

\usepackage{mathptmx}       % selects Times Roman as basic font
\usepackage{helvet}         % selects Helvetica as sans-serif font
\usepackage{courier}        % selects Courier as typewriter font
\usepackage{type1cm}        % activate if the above 3 fonts are
\usepackage{makeidx}         % allows index generation
\usepackage{graphicx}        % standard LaTeX graphics tool
\usepackage{multicol}        % used for the two-column index
\usepackage[bottom]{footmisc}% places footnotes at page bottom

\def\Esp#1{\mathbb{E}\left[#1\right]}

\usepackage{amssymb}
\usepackage{algorithm}
\usepackage{algorithmic}  
\usepackage{subfigure}

\makeindex             % used for the subject index
\begin{document}

\title*{Hedging Swing contract on gas market}
\author{Warin X.}
\institute{ Xavier Warin  \at Xavier Warin, \small{EDF R\&D \& FiME}, Laboratoire de Finance des March\'es de l'Energie (www.fime-lab.org)  \email{xavier.warin@edf.fr}}
%

\maketitle

\abstract*{
The aim of this paper is to give an efficient algorithm to follow the dynamic hedge of a Swing contract on the gas market.
Methodology is given, and a numerical test based on a real contract shows that an efficient hedge has to  deal with the components
 involved in the index.}

\section{Introduction}
Swing options on the gas market are american style option where daily quantities exercices are constrained and global quantities exerciced each year constrained too. 
The option holder has to decide each day how much he consumes of the quantities satisfying the constraints and tries to use a strategy in order to maximize its 
expected profit.
Based on the indexation principle, the pay off fonction is a spread between the spot gas market and the value of an index composed of the past average of some commodities' spot or future
prices. Most of the time the commodities involved are gas and oil.\\
When the index is a fixed one (then it is a strike), the valuation of Swing option is a classical problem solved by dynamic programming. In order to implement
the dynamic programming method , first trinomial trees were used  \cite{jaillet1} to estimate conditional expectation, then Longstaff Schwartz method \cite{long2001} using regression and Partial Differential Equation methods \cite{wihlem}.\\
At the opposite, Swing options on the gas market are very high dimensional problems due to the index definition.
This problem can be linked to the problem of pricing  moving average american options. These moving averaging 
options with early exercice features have been studing in \cite{broadie}, \cite{Dai}, \cite{grau}. A common approach (see \cite{broadie})
is to use least square Monte Carlo to estimate conditional expectation in Montecarlo algorithm. Recently \cite{bernhart} derived a new
methodology based on exponentially weighted Laguerre polynomials expansion to approximate the moving average processes. The methodology developped is efficient but numerical results
showed that the usual regression method regressing on both the gas price and the index value is accurate enough.\\
If the question of the valuation of these contracts is rather well explored, the hedge of such a contract has never been studied.
Most of the time practionners just hedge the gaz spot component in the index ignoring the stochasticity of the index.  Besides when practionners
want to simulate the portfolio risk they have to include the dynamic hedge they plan for next years in order to assess their risk accurately.
As for gas storage, a methodology based on tangent process has been developped by \cite{warin} to evaluate the Delta and follow the hedging strategy. The mathematical proof of this approach based on the Danskin's theorem is given in \cite{bonnans}.
In this paper the methodology based on the tangent forward method calculated during a dynamic programming algorithm is developped for the gas Swing options. Its accuracy is tested on a realistic contract  where:
\begin{itemize}
\item prices  follow a two factor model (see for example \cite{schwarz}),
\item hedging products available are day ahead products till the end of the week, week ahead products till the end of the month, and monthly contract as they can be found at Henry Hub.
\end{itemize}
The efficiency of the hedge and the accuracy of the calculations are  studied depending on :
\begin{itemize}
\item the regressor used to calculate conditional expectation,
\item the products involved in the index  that we  decide to hedge,
\item the frequency of the hedge with respect to  the  products involved in the index including exchange rates.
\end{itemize}
In the sequel we suppose that conditional expectation are calculated by the Longstaff-Schwartz method \cite{long2001} adapted with local functions 
as explained in \cite{bouchardwarin}.\\
In the next section \ref{secII}, the contract and the price model  are  described. In the following section \ref{secIII}, the methodology used to simulate the dynamic hedging
is explained. Next numerical results are given and a concluding section \ref{secVI} gives some recommandations in particular in terms of frequency of the
dynamic hedging and the components to hedge.

\section{The contract and the price modelization}
\label{secII}
\subsection{The contract}
\begin{figure}[ht]
\centering
\includegraphics[width=12cm]{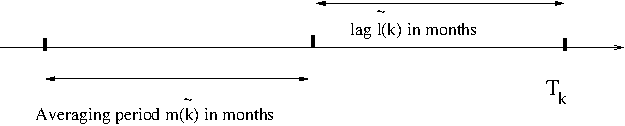}
%
%\vspace*{-0.25cm}
%
\caption{\label{indexfig} Lag and averaging windows for $\hat{S}^{\tilde l(k),\tilde m(k)}_{T_{k(i)}}$ calculation }
\end{figure}

The Swing contract is first characterized by the quantities that can be purshased. Daily exercice dates are given in days from $t_{start}$ to $t_{end}$.
Each day $t$ the option holder can exercise a quantity $q_t$ satisfying $$ 0 \le q_t \le q_{max}$$
Moreover  the global quantity consummed  $Q_t$ is constrained $$ Q_{min} \le  \sum_{t_{start} \le t \le t_{end}} q_{t} \le Q_{max}.$$
The unitary pay off (for a quantity one) is given at a day $t$ by $$S_t^0- I_t$$
where $S_t^0$ is the gas price and $I_t$ an index constant on each interval $[T_i,T_{i+1}[$ where $T_i$ ($i=0$ to $\hat n$) are update dates defined in the contract.
These dates correspond to the beginning of some months and typical values for $T_i$ are first day of each month, or first day of every two or three months.\\
For each  exercice day $i$, we note $k(i)= \max\{j, T_j \le i \}$.
In the case of an additive index, the structure of the index  for each exercice day $i$ is defined by the affine sum of $N$ components by :
$$
I_{i} = I_{T_{k(i)}} = a_0 + \sum_{j=1}^N a_j ( {\hat{S^j}}^{ \tilde l(k(i)),\tilde m(k(i))}_{T_{k(i)}} - b_j) X^j_{T_{k(i)}}
$$
where :
\begin{itemize}
\item $\hat{S^j}^{\tilde l,\tilde m}_{T} =  \frac{1}{\tilde m} \sum_{q=1}^{\tilde m} S^j_{T - \tilde l-q}$ is the average of commodity $j$ (potentially a forward price) on 
the $\tilde m$ months expressed in days  preceding the lag corresponding to $\tilde l$ months  expressed in days involved in the index valid from date 
$T$ (see figure \ref{indexfig}),
\item $X^j_{T_{k}}$ is the exchange rate between the domestic currency and the foreign currency of the commodity $j$ at date $T_{k}$,
\item $a_j$, $j=0 ... N$, $b_j =1, ... N$ are coefficients given by the contract.
\end{itemize}
\begin{remark}
In some simple cases the lag and averaging windows are the same for each component of the index and the contract windows can be decribed by the triplet $m,l,p$
where $p$ gives in months the validity period for the index, $l$ the lag periode and $m$ the averaging window length. 
\end{remark}
\begin{remark}
In the formula, the exchange rates involved  could be imposed by the  contract at date $T(k)-\tilde l(k)$ or an average of the exchange rate could be imposed. 
\end{remark}
\subsection{Price Models}
All over this section, we shall  consider a $\tilde{n}$-dimensional Brownian motion $\left \{ z^1,\ldots,z^{\tilde{n}}\right \}$ with correlation matrix $\rho$ on a probability space $(\Omega,{\cal F},{ \mathbb P})$ endowed with the natural (completed and right-continuous) filtration ${ \mathbb F}=({\cal F}_t)_{t\le T}$ generated by   $\left \{ z^1,\ldots,z^{\tilde{n}} \right \}$  up to some fixed time horizon $T>0$. The $\tilde n$ value will depend on the different models
 used for commodity prices and exchange rates.
\subsubsection{ Future Price Model} 
We suppose that the uncertainties in the commodity prices $S^j_t$, $j=0,N$ follows under the risk neutral measure a $\tilde{n}$-dimensional Ornstein-Uhlenbeck process. 
The following SDE describes our
uncertainty model for the forward curve $F^j(t,T)$ giving the prices of a unitary amount of a commodity $j$ at day $t$ for delivery at date $T$  ( see \cite{clewlow}):
\begin{equation}
\frac{dF^j(t,T)}{F^j(t,T)} =  \sum_{i=1}^n \sigma_i^j(t) e^{-a_i^j(T-t)} dz^{j n+i}_t, \\
\label{eq:alea}
\end{equation}
with $\sigma_i^j$ some
volatility parameters and $a_i^j$ mean reverting parameters for commodity $j$. 
\begin{remark}
Most of the time a two factors model is used. In this model,
the first Brownian motion describes swift changes in the future curve, the second one describes structural changes in the gas market and deals with long term changes in the curve. The mean reverting parameter is generally taken equal to zero for the second term.
\end{remark}
With the following notations:
\begin{eqnarray}
 V^j(t_1,t_2) & =  &\displaystyle{ \int_{0}^{t_1} } \left\{ \sum_{i=1}^n \sigma_i^j(u)^2 e^{-2 a_i^j (t_2-u)} +  \right. \nonumber\\
 & &  \left . 2 \sum_{i=1}^n \sum_{k=i+1}^n \rho_{i+jn,k+jn} \sigma_i^j(u)
 e^{- a_i^j (t_2-u)} \sigma_k^j(u)  e^{- a_k^j (t_2-u)}  \right \} du , \nonumber \\
 W^{i,j}_t  & = &  \displaystyle{\int_{0}^t} \sigma_i^j(u) e^{-a_i^j (t-u)}  dz^{i+jn}_u ,  \mbox{ for } i =1,\ldots, n \nonumber
 \end{eqnarray}
the integration of equation (\ref{eq:alea})  gives :
\begin{equation}
F^j(t,T)  =  F^j(t_0,T) e^{  -\displaystyle{\frac{1}{2}} V^j(t,T)+  \sum_{i=1}^n e^{-a_i^j (T-t)} W^{i,j}_t }.
\end{equation}
With this modelization, the spot price for commodity $j$ is defined as the limit of the future price :
\begin{equation}
S^j_t =  \lim_{T \downarrow t} F^j(t,T).
\end{equation}
As for the instantaneous exchange rate for commodity $j$, $X^j_t$, we choose a simple model
\begin{equation}
\frac{dX^j_t}{X_t} = (r_d(t) - r_f^j(t)) dt + \sigma_X^j dz^{nN+j}_t,
\end{equation}
for $j=1,..,N$ where
\begin{itemize}
\item $r^j_f(t)$ is the short rate for the currency of commodity $j$,
\item $r_d(t)$ is the domestic short rate,
\item $\sigma_X^j$ is the volatility of the foreign currency associated to commodity $j$ in domestic numeraire.
\end{itemize}
In the sequel we suppose that the domestic interest rate $r_d(t)$ is 0 for simplicity.
\subsection{Objective function}
Under the martingale probability, the owner of the option will try to optimize it's profits by maximizing the following  
\begin{eqnarray}
  J(q)& =  &  \Esp{ \sum_{t=t_{start}}^{t_{end}} \left\{ q_t (S_t^0 -I_t) \right\}} 
\label{progdyn}
\end{eqnarray}
where $q= \{ q_t , t \in [t_{start}, ..., t_{end}] \}$,
under the constraints  : 
\begin{equation}
  \left \{
\begin{array}{ccc}
 q_t & \in & [0,q_{max}] \\
 Q_{min}                   \le & \sum_{t=t_{start}}^{t_{end}} q_t &  \le Q_{max} \\
\end{array}
\right.
\label{constraints}
\end{equation}
We introduce $$Q_t =  \sum_{t=t_{start}}^{t-1} q_t$$, $${\cal U}=  \{ (q_{t_{start}}, ..., q_{t_{end}}) , q_t \in {\cal F}_t  ,  0 \le q_t \le q_{max} , Q_{min} \le Q_{t_{end}} \le Q_{max} \},$$

%and  $$Z_t = \left\{ Z^_{\tilde t}, \tilde t \le t, j \le N \}, \{X^j_{\tilde t},\tilde t \le t, j \le N \right\}$$ the stochastic part of
and  $$Z_t = \left\{ z^j_{\tilde t}, \tilde t \le t, j \le \tilde{n}= 2(N+1)n + N \right\}$$ the stochastic part of the state vecteur at date $t$.

The solution of the problem is given by $J^*$ :
\begin{eqnarray}
J^* =        \max_{q \in{\cal U}} J(q).
\label{OptControl}
\end{eqnarray}
According to the  Bellman principle, the optimal function value at date $t$  expressed as a function of $Q_t$ and $Z_t$ satisfy :
\begin{equation}
J^*(t,z,Q_t) = \sup_{ \begin{array}{l}
                      0 \le q \le q_{max},\\
                      Q_t + q+\sum_{tt> t} q_{max} \ge Q_{min},\\
                      Q_t + q \le Q_{max}
\end{array}} \left\{ q (S^0_t- I_t) + \Esp{ J^*(t+1,Z_{t+1},Q_{t}+q) ~|~ Z_t=z } \right\} 
\label{progdyn}
\end{equation}
This approach is a classical one which has been used for example recently in \cite{wihlem}.

\section{Optimization and dynamic hedging}
\label{secIII}
In the first part we give the equation for the dynamic hedge of each commodity and echange rate of the problem according to the methodology developped in \cite{warin}.
We then explain why Delta have to be approximated and aggregated to be calculted on realistic problems.
\subsection{Continuous version of the hedge}
During the dynamic programming resolution,  the daily quantities to exercice can be  discretized if necessary or a bang bang approximation (which can be exact  \cite{bardou}) can be used.
We note $q^*(t,Z_t,Q_t) $ the optimal volume exerciced at date $t$ when the quantity already consumed is $Q_t$.
We introduce $Q^{*,i}_{ii}(z,c)$ the optimal volume level at date $ii$ starting at level $c$ at date $i$ following a trajectory $Z_t$ with  $Z_i= z$.
The optimal volume is ${\cal F}_{i}$-mesurable  and follows
\begin{eqnarray}
Q^{*,i}_{ii}(z,c) &= & c, \forall ii \le t_{start}, ii \ge i,  \nonumber \\
Q^{*,i}_{ii}(z,c) & = &  c + \sum_{k=t_{start} \vee i }^{ii-1} q^*(k,Z_{k},Q^{*,i}_{k}(z,c)) \mbox{ for }  t_{start} \le ii \le t_{end}. 
\label{optConso}
\end{eqnarray}
Thus $\sum_{k=t_{start}}^{t_{end}} q^*(k,Z_{k},Q^{*,0}_{k}(z,0))$ corresponds to the sum of the optimal volumes exercised following the optimal strategy starting with a zero consummed volume at date $0$ with  $Z_{0}=z$.\\
Following the methodology  in \cite{warin} we introduce  the forward tangent process for commodity $j$ noted $Y_t^{j,T}$ satisfying :
\begin{equation}
Y_t^{j,T}   =   e^{ -\frac{1}{2} V^j(t,T) + \sum_{i=1}^n e^{-a_i^j (T-t)} W^{i,j}_t },  j=0,...,N,
\label{tangentForward}
\end{equation}
and for the  exchange rate $j$ the classical tangent process \cite{gobetgreeks} :
\begin{equation}
Y_{t}^j =  e^{-\int_{0}^t  r_f^j(s)) ds  -\frac{1}{2} (\sigma_X^j)^2 +\sigma_X^j dz^{nN+j}_t},  j =1,...,N.
\label{tangentX}
\end{equation}
As prooved in \cite{bonnans}, introducing the ${\cal F}_{m}$ random variable :
\begin{equation}
 D^0(i,m,z,Q)  =   q^*(m,Z_{m} ,Q^{*,i}_m(z,Q)) Y_{m}^{0,m}.
\label{ProcessTang}
\end{equation}
the sensibility of the contract value with respect to the arbitrage market (spot gas) for delivery at date $m$ is then given by
\begin{eqnarray}
\label{CDelta}
\Delta^0(i,m,z,Q)=  \mathbb E[D^0(i,m,z,Q)~|~ Z_{i}=z]/Y_{i}^{0,m} , t_{end} \ge m > i
\end{eqnarray}
For the component $j$ of the index, introducing
\begin{eqnarray}
D^j(i,m,l,z,Q)  =   q^*(l,Z_{l} ,Q^{*,i}_l(z,Q)) Y_{m}^{j,m} ,
\label{eqSup}
\end{eqnarray}
and for $m \ge i$
\begin{eqnarray}
{\bar{D}}^j(i,m,z,Q) =  a_j \sum_{k} \sum_{l \in [T(k),T(k+1)[} D^j(i,m,l,z,Q) X^j_{T(k)}  \frac{1_{T(k) -  \tilde l(k) > m > T(k) - \tilde l(k)- \tilde m(k)}}{\tilde m(k)}, 
\label{DeltaBarCommo}
\end{eqnarray}
 the sensibility of component $j$ of the index at date $i$ and delivery at date $m$ ($m > i$) is then :
\begin{eqnarray}
\Delta^j(i,m,z,Q)= -  \mathbb E[ {\bar{D}}^j(i,m,z,Q) ~|~ Z_{i}=z]/Y_{i}^{j,m} 
\label{compoSen}
\end{eqnarray}
\begin{figure}[ht]
\centering
\includegraphics[width=12cm]{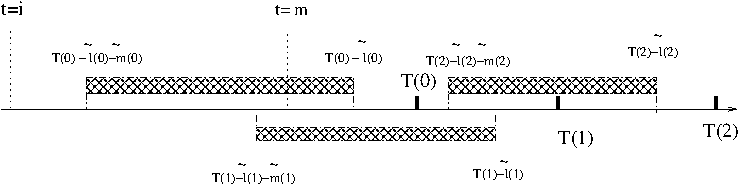}
%
%\vspace*{-0.25cm}
%
\caption{\label{figSensibilite} Example of sensibility for first component of the index}
\end{figure}
To ease  the comprehension of formula (\ref{compoSen}), we explain  it for the first index component (commodity one) on figure \ref{figSensibilite} when the index is only updated tree times
at date $T_0$, $T_1$ and $T_2$. At date $i$, the first component of the index corresponding to commodity 1 with delivery at date $m$ has a direct impact on the indexed price at which the gas will be paid  for all the dates between $T_0$ and $T_2$ so the sensibility can be written as the sum of a component  for delivery during the first and the second period. $\displaystyle{\sum_{l \in [T(0),T(1)[}} q^*(l,Z_{l} ,Q^{*,j}_l(z,Q))$ for example is the optimal consumption on first periode exerciced with the index $I_{T(0)}$ constant on first period and thus the sensibility in equation (\ref{compoSen}) becomes :
\begin{eqnarray}
\Delta^1(i,m,z,Q)=   -\frac{a_1}{Y_{i}^{1,m}} &  \left ( \frac{1}{\tilde m(0)} \mathbb E[ \displaystyle{\sum_{l \in [T(0),T(1)[}} q^*(l,Z_{l} ,Q^{*,i}_l(z,Q)) X^1_{T(0)}  Y_{m}^{1,m} ~|~ Z_{i}=z]  + \right. \nonumber\\
&  \left. \frac{1}{\tilde m(1)}  \mathbb E[ \displaystyle{\sum_{l \in [T(1),T(2)[}}   q^*(l,Z_{l} ,Q^{*,i}_l(z,Q)) X^1_{T(1)}  Y_{m}^{1,m}~|~ Z_{i}=z] \right )  \nonumber
\end{eqnarray}
As for the exchange rate corresponding to component $j$ of the index, introducing
$$
D^j_X(i,m,l,z,Q)  =   q^*(l,Z_{l} ,Q^{*,i}_l(z,Q)) Y_{m}^{j} ,
$$ 
and 
$$
\bar{D}^j_X(i,z,Q)  =   a_j \sum_{k,T(k)>i\quad} \sum_{l \in [T(k),T(k+1)[}  D^j_X(i,T(k),l,z,Q) \hat{S^j}^{\tilde l(k),\tilde m(k)}_{T(k)} ,
$$
the sensibility is given by
\begin{eqnarray}
\Delta^j_X(i,z,Q)= - \mathbb E[ \bar{D}^j_X(i,z,Q) ~|~ Z_{i}=z]/Y_{i}^{j} 
\label{sensExRate}
\end{eqnarray}
Here for each period $k$ that has not began, we assess the optimal volume exerciced on the period multiplied by the tangent process at $T(k)$ for the exchange rate 
multiplied by the $a_j$ coefficient and the average value of the commodity $j$ used for the index on period $[T(k),T(k+1)[$.  The results are obtained by summing
on the periods and taking conditional expectation.

\subsection{Discretized algorithm}
In order to solve equation (\ref{progdyn}),  the  methodology in \cite{bouchardwarin} derived from the Longstaff-Schwartz methodology \cite{long2001} 
can be used as in the case of American options.
Using this methodology adapted to Swing option, optimal exercices and corresponding hedging strategies calculated at a date $t$ depend on quantities consumed before this date.
So optimal values and hedge have to be calculated on a grid used to discretize the possible volume consumed. Interpolation between Bellman values and hedging strategies
is then used during the dynamic programming process.\\
The volume is discretized on the grid   $$ Q_l = l \Delta  ,\quad   l= 0 ,\ldots,  L =Q_{max}/\Delta $$ where $\Delta$ is the mesh size.
Similarly to the case of American option, a Monte Carlo method is used to  get some  prices simulations $(S_i^{j})^{\tilde j}$ for  $ \tilde j= 1, \ldots, M$ at day $i=0$ to $t_{end}$, for $j=0, ..., N$. In the sequel $(.)^{\tilde j}$ will stand for the simulation $\tilde j$ of random variable $(.)$.
 The conditional expectation $\mathbb E[  ~|~ Z_t]$ has to be calculated in very high dimension but
as shown in \cite{bernhart},  it is possible to approximate the operator $\mathbb E[  ~|~ Z_{i}]$ by the operator
 $\mathbb E[  ~|~ S^0_i, I_i]$ with accuracy leading to regression in dimension 2.
 In the numerical part of the article a small study on the regressor will be achieved  more thoroughly.
In the sequel we note $\hat \mathbb E[  ~|~ \hat Z_{t}]$ an approximation of the conditional expectation operator $ \mathbb E[  ~|~ Z_{t}]$ obtained by regression
with \cite{bouchardwarin} methodology where the information contained in the stochastic state vector $Z_t$ has been approximated by some information contained in  $\hat Z_t$ vector.
\\

We note $\hat q^*$ and $\hat Q^*$  the estimation of the optimal volume $q^*$ and optimal volume levels $Q^*$ obtained by  the Longstaff-Schwartz method.
We  note $\hat D^0$, $\hat D^j$,$\hat {\bar D}^j$,
$\hat D^j_X$, $\hat {\bar D}^j_X$, $j=1,N$  the estimations of the variable  $D^0$, $D^j$,${\bar D}^j$, $D^j_X$ and ${\bar D}^j_X$ on Montecarlo simulation for optimal control calculated $\hat q^*$ .\\
As explained in \cite{warin}, knowing $ \hat D^0(i+1,m,(\hat Z_{i+1})^{\tilde j},Q_l)$  for $m = i+1, \ldots t_{end}-1$, $l=1,\ldots L$, $ \tilde j=1, \ldots M$
and using the equality
$$ \hat Q^{*,i}_m((\hat Z_{i})^{\tilde j},Q) = \hat Q^{*,i+1}_m((\hat Z_{i+1})^{\tilde j},Q + \hat q^*(i,(\hat Z_{i})^{\tilde j},Q))$$ for $t_{end} > m> i$,
the  following backward recursion can be used to calculate $\hat D^0(i,j,Z_i,Q)$ at day $i$ for all $j \ge i$, for all $Q \in \{Q_l / l=0,...,L \}$:
\begin{equation}
\left \{ \begin{array}{lll}
\hat D(i,i,(\hat Z_{i})^{\tilde j},Q) & =&  \hat q^*(i,(\hat Z_{i})^{\tilde j},Q) (Y_{i}^{0,j})^{\tilde j} \\
\hat D(i,m,(\hat Z_{i})^{\tilde j},Q) & = & \hat D(i+1,m,(\hat Z_{i+1})^{\tilde j},Q+\hat q^*(i,(\hat Z_{i})^{\tilde j},Q)),   m = i+1, t_{end}
\end{array}
\right.
\label{DeltaRec}
\end{equation}
As for as the current backward recursion on option value, an interpolation on $\hat D$ values is necessary to estimate $\hat D(i+1,.,.)$ for the
stock point $Q+\hat q^*(i,(\hat Z_{i})^{\tilde j},Q)$.\\
The Longstaff-Schwartz estimator of the conditional Delta is then evaluated for $m > k$ by
\begin{eqnarray}
\label{CDeltaDis}
\hat \Delta^0(i,m,(\hat Z_i)^{\tilde j},Q)= \hat { \mathbb E}[\hat D(i,m,\hat Z_{i},Q)~|~ \hat Z_{i}=(\hat Z_i)^{\tilde j}]/(Y_{i}^{0,m})^{\tilde j} .
\end{eqnarray}

Similar recursion can be used for  $\hat D^j$ with $j>0$. We first give the $\hat D^j$ approximation of $D^j$  :
\begin{equation}
\hat D^j(i,m,l,(\hat Z_i)^{\tilde j},Q)  =   q^*(l,(\hat Z_l)^{\tilde j},Q^{*,i}_l((\hat Z_i)^{\tilde j},Q)) (Y_{m}^{j,m})^{\tilde j}  , i \le m <l <t_{end} 
\label{initIndex}
\end{equation}
Introducing  $\hat {\bar D}^j$ the ${\bar D}^j$ approximation :
$$ \hat {\bar D}^j(i,m,(\hat Z_i)^{\tilde j},Q) =  a_j \sum_{k}\sum_{l \in [T(k),T(k+1)[} D^j(i,m,l,(\hat Z_i)^{\tilde j},Q) \frac{1_{T(k) - \tilde l(k) > m > T(k) - \tilde l(k)- \tilde m(k)}}{\tilde m(k)}$$
we then calculate the $\hat {\bar D}^i$ values during the recursion easily supposing that $\hat {\bar D}^j(i+1,j,(\hat Z_j)^{\tilde j},Q)$ have been previously calculated with
\begin{equation}
\left \{ \begin{array}{lll}
\hat {\bar D}^j(i,i,(\hat Z_i)^{\tilde j},Q) & =&  a_j \sum_{k}\sum_{l \in [T(k),T(k+1)[} D^j(i,i,l,(\hat Z_i)^{\tilde j},Q) X^j_{T(k)} \frac{1_{T(k) - \tilde l(k) > i > T(k) - \tilde l(k)- \tilde m(k)}}{\tilde m(k)} \\
\hat {\bar D}^j(i,m,(\hat Z_i)^{\tilde j},Q)&  = & \hat {\bar D}^j(i+1,m,(\hat Z_{i+1})^{\tilde j},Q+\hat q^*(i,(\hat Z_{i})^{\tilde j}),Q), i < m
\end{array}
\right.
\label{DeltaRecIndex}
\end{equation}
Equation (\ref{DeltaRecIndex}) allows  to calculate the hedge :
\begin{eqnarray}
\hat \Delta^j(i,m,(\hat Z_i)^{\tilde j},Q)= -  \mathbb E[\hat {\bar D}^j(i,m,l,\hat Z_i,Q) ~|~ \hat Z_{i}=(\hat Z_i)^{\tilde j}]/(Y_{i}^{j,m})^{\tilde j}
\label{compoSenbDis}
\end{eqnarray}
As for the exchange rate we introduce $\hat D^j_X$ the $D^j_X$ approximation :
\begin{equation}
\hat D^j_X(i,m,l,(\hat Z_i)^{\tilde j},Q)  =   q^*(l,(\hat Z_l)^{\tilde j},Q^{*,i}_l((\hat Z_i)^{\tilde j},Q)) (Y_{m}^{j})^{\tilde j} , i < m <l 
\label{initIndex}
\end{equation}
and  the $\hat {\bar D}^j_X$ approximation of $ {\bar D}^j_X$
\begin{equation}
\hat  {\bar D}^j_X(i,(\hat Z_i)^{\tilde j},Q) =  a_j \sum_{k/T(k)>i\quad} \sum_{l \in [T(k),T(k+1)[}  \hat D^j_X(i,T(k),l,(\hat Z_i)^{\tilde j},Q) (\hat{S^j}^{\tilde l(k),\tilde m(k)}_{T(k)})^{\tilde j}
\end{equation}
The following recursion adding at each day all the contribution on the sensibility due to the following  day can be used during the backward recursion :
\begin{equation}
\left \{ \begin{array}{lll}
\hat  {\bar D}^j_X(i,(\hat Z_i)^{\tilde j},Q) &= & 0 , j \ge T(n),  \\
\hat  {\bar D}^j_X(i,(\hat Z_i)^{\tilde j},Q) &= & a_j \sum_{k} \delta_{T(k)}(i+1) \sum_{l \in [T(k),T(k+1)[}  \hat D^j_X(i,T(k),l,(\hat Z_i)^{\tilde j},Q) (\hat{S^j}^{\tilde l(k),\tilde m(k)}_{T(k)})^{\tilde j}  \\
& & + \hat {\bar D}^j_X(i+1,(\hat Z_{i+1})^{\tilde j},Q+\hat q^*(i,(\hat Z_{i})^{\tilde j},Q)) , i < T(n)
\end{array}
\right.
\label{ExRateRecur}
\end{equation}
leading to the Delta approximation
\begin{eqnarray}
\hat \Delta^j_X(i,(\hat Z_i)^{\tilde j},Q)= - \mathbb E[ {\bar D}^j_X(i,\hat Z_i,Q) ~|~ \hat Z_{i}= (\hat Z_{i})^{\tilde j}]/(Y_{i}^{j})^{\tilde j} 
\label{sensExRateDis}
\end{eqnarray}
Using equation (\ref{DeltaRec}), (\ref{CDeltaDis}) sensibility with respect to gas is thus calcultated, using equation (\ref{DeltaRecIndex}) (\ref{compoSenbDis})
sensibility with respect to commodity index components calculated and using equation (\ref{ExRateRecur}) and (\ref{sensExRateDis}) sensibility
to exchange rate calculated.

\subsection{Aggregation}
Not all product are available on the market. We note ${\cal Q}_j^i$ the set of future products available at date $i$ for commodity $j$, and for all $p \in {\cal Q}_i^j$,
${\cal P}_p^j$ the delivery period associated to product $p$ available for commodity $j$, ${\eta}_p^j$ the beginning of the delivery period.
Supposing that $ \forall t > 0, \forall p \in {\cal Q}_i^j, \forall \tilde{i} > i$ there exist ${\cal Q}^{j,p} \subset {\cal Q}_{\tilde i}^j$ such that 
$ {\cal P}_p^j  = \cup_{\tilde{p} \in {\cal Q}^{j,p}} {\cal P}_{\tilde p}^j$ then it is possible to aggregate an approximated conditional Delta at date $i$ per product with an ad hoc rule so that a dynamic programming approach is still usable.
 A first way to get the Delta on a delivery period $p$ is to average the Deltas calculated with equations (\ref{CDeltaDis}) and (\ref{compoSenbDis})
 calculated during recursion on the delivery period.
This approach is memory cumbersome as noted in \cite{warin}.
It is far more efficient to calculated the Delta directly on the product $p$ during the backward recursion.
\cite{warin} proposed to use in the continuous framework :
\begin{equation}
\left \{ \begin{array}{lll}
\tilde D^0(i,p,Z_i,Q )  & = &   \mathbb E[ \sum_{m \in  {\cal P}_p^0} \hat q^*(m,Z_{m},\hat Q^{*,i}_m(Z_{i},Q)) Y_{m}^{0,m} F^0(0,t_m)~|~ Z_{i}] , p \in {\cal Q}_j^0  \nonumber\\
\tilde \Delta^0(i,p,Z_i,Q ) & =  & {\tilde D}^0(i,p,Z_i,Q ) / ( Y_{i}^{\eta_p^0} \sum_{m \in {\cal P}_p^0} F^0(0,t_m))
\end{array}
\right.
\label{DeltaCond}
\end{equation}
where $\tilde \Delta^0(i,p,Z_i,Q)$ represents the power to invest at date $i$ for product $p$ for a gas volume  level consumed $Q$ and a stochastic state vector $z$. 
As shown by  \cite{warin}, the $ \tilde D^0(i,p,z,Q)$ can be calculated during the dynamic programming recursion using :
\begin{eqnarray} 
     \tilde D^0(i,p,Z_i,Q)     & = &   \hat { \mathbb E}[1_{ i+1 \in{\cal P}_p} \hat q^*(i+1,Z_{i+1},\hat Q^{*,i}_{i+1}(Z_{i},Q)) Y_{i+1}^{0,i+1} F^0(0,i+1)~|~ Z_{i}] \nonumber     \\
          &  & + \sum_{ \tilde p \in {\cal Q}^{0,p}} \hat { \mathbb E}[ \tilde D(i+1,\tilde p,Z_{i+1}, Q +\hat q^*(i,Z_{i},Q) )~|~ Z_{i}=z]. \nonumber   
\end{eqnarray}
The same procedure can be applied  to derive a Delta estimation for the index commodity component $j>0$ :
\begin{equation}
\left \{ \begin{array}{lll}
 \tilde D^j(i,p,Z_i,Q)  & = &   \mathbb E[ \sum_{m \in  {\cal P}_p^j} {\bar D}^j(i,m,Z_i,Q) F^j(0,t_m)~|~ Z_{i}] , p \in {\cal Q}_i^j \\
\tilde \Delta^j(i,p,Z_i,Q) & =  & \tilde D^j(i,p,Z_i,Q) / ( Y_{i}^{\eta_p^j} \sum_{m \in {\cal P}_p^j} F^j(0,t_m))
\end{array}
\right.
\label{DeltaCondCommo}
\end{equation}
Using equation (\ref{DeltaBarCommo})  we get
\begin{eqnarray}
\tilde D^j(i,p,Z_i,Q)   = \sum_{m \in  {\cal P}_p^j} a_j \sum_{k} \sum_{l \in [T(k),T(k+1)[} F^j(0,t_m) \frac{1_{T(k) -  \tilde l(k) > m > T(k) - \tilde l(k)- \tilde m(k)}}{\tilde m(k)}
 \mathbb E[ D^j(i,m,l,Z_i,Q)~|~ Z_{i}] \nonumber
\end{eqnarray}
Besides from equation (\ref{eqSup})
\begin{eqnarray}
\mathbb E[ D^j(i,m,l,Z_i,Q)~|~ Z_{i+1}] & = & \mathbb E[ q^*(l,Z_{l} ,Q^{*,i}_l(Z_i,Q)) Y_{m}^{j,m}~|~ Z_{i+1} ],   \mbox{for} m > i+1 \nonumber \\
& = & \mathbb E[ q^*(l,Z_{l} ,Q^{*,i+1}_l(Z_{i+1},Q+q^*(Z_i,Q))) Y_{m}^{j,m} ~|~ Z_{i+1} ] \nonumber  \\
& = & \mathbb E[D^j(i,m,l,Z_{i+1},Q+q^*(Z_i,Q)~|~ Z_{i+1}] \nonumber
\end{eqnarray}
so that
\begin{eqnarray}
\mathbb E[ D^j(i,m,l,Z_i,Q)~|~ Z_{i}] & = & \mathbb E[D^j(i,m,l,Z_{i+1},Q+q^*(Z_i,Q)~|~ Z_{i}] \nonumber
\end{eqnarray}
and
\begin{eqnarray}
\tilde D^j(i,p,Z_i,Q)   & = & \hat { \mathbb E}[1_{ i+1 \in{\cal P}_p^j} {\bar D}^j(i,i+1,l,Z_i,Q))~|~ Z_{i}] + \nonumber \\
& & \sum_{ \tilde p \in {\cal Q}^{j,p}} \hat { \mathbb E}[  {\tilde D}^j(i+1,p,Z_{i+1},Q+q^*(Z_i,Q))~|~ Z_{i}] 
\label{DeltaAppCommo}
\end{eqnarray}
which can be easily calculated during backward recursion.

\section{Numerical results}

In first subsection we present the contract used for this study and we give the different numerical parameters used in calculation :
\begin{enumerate}
\item the  number of basis function used for each dimension for regression used to estimate conditional expectation \cite{bouchardwarin},
\item the number of particules used in an optimization part calculating the optimal value and the conditional hedge store in files,
\item the number of particules used in a simulation part where the dynamic hedge is carried out.
\end{enumerate}
In a second subsection a study on the regression dimension is achieved and in a third subsection we test for the contract the sway of each component on
the hedge's efficiency.
At last we test the effect of the  hedge frequency on the hedge efficiency. 
\subsection{Contract description and general parameter} 
We take the following example derived from a real Swing contract.
We suppose that the Swing is contracted for year 2007 and that the day of valorisation is 1th of  april 2006.
The flexibility is such that each day normalized quantities are such that $q_{max}=0.4$, $Q_{min}=91$,$Q_{max}=146$.
\begin{remark}
With this kind of parameters the $Q_{max}$ constraint is ineffective.
\end{remark}
The  arbitrage market taken for $S^0$ is the spot  gas at Zeebruge hub. 
The index is composed of two components :
\begin{itemize}
\item The first component is described by parameters $(9,0,1)$ such that the average is taken on 9 months ($\tilde m$ equivalent to 9 months),
with no lag and the index component is changed every month. The first commodity is the brent.
\item The second component of the index is described by parameters
$(1,0,1)$ meaning that the average is taken during 1 month ($\tilde m$ equivalent to 1 month) , no lag ($\tilde l =0$)
and the index component 2 is changed every month. The second commodity is  the month ahead gas  on TTF market.
\end{itemize}
The parameters $a_i, b_i$ have been adapted such that the  Swing is at the money (the corresponding swap contract has a zero value) :
$a_0 =2.2$, $a_1=0.1757$, $a_2=0.2714$, $b_1=45.16$, $b_2=0$.\\

The model for each commodity are described by a two factor model ($n=2$)  :
\begin{itemize}
\item The Brent annual short term volatility  is equal to $0.28$, the annual short term
mean reverting  is equal to $0.1$, and the annual long term volatility is equal to $0.1$, the quotation is in dollar.
\item As for the spot/future gas market, the short term annual volatility is equal to $1$, the annual short term mean reverting equal to $80$, while the
long term volatility is equal to $0.1$. The quotation is in Euro the
domestic market.
\item Spread between the euro and dollar interest rate is 0, and the annual volatility between the two currency is $0.11$.
\end{itemize}
The daily forward curves are generated by Monte Carlo and at each day, depending on the market product available, this curves is averaged on each delivery period
associated to each product giving products values that will be used during hedging.\\

The parameter used for the calculations are the following :
\begin{itemize}
\item we use 80000 trajectories in optimization to calculate the Swing value and the hedge,
\item when regressing in dimension 1, $6$ basis functions are taken. When regressing in dimension 2, $6 \times 4$ basis function are taken, while when
regressing in dimension 3, $6 \times 4 \times 1$ basis functions are taken.
\item The global quantities $Q_l$ are discretized with a step of $0.4$ and the local quantities with a step $0.1$.
\end{itemize}
Due to the calculation time  and most of all due to the memory needed on computer, the Delta calculations have been parallelized by mpi (http://www.mcs.anl.gov/research/projects/mpi/) on a cluster during
optimization and parallelization on scenarios has been achieved in simulation.
A single optimization takes roughly 5 hours on 12 cores of a cluster and half an hour in simulation.
  
\subsection{Regression tests}
In the section we test three regressors :
\begin{itemize}
\item for the first one, we approximate $Z_t$ by $S^0_t$,
\item for the second one, we approximate $Z_t$ by $(S^0_t,I_t)$,
\item for the last one, we approximate  $Z_t$ by $(S^0_t,I_t, \tilde I_t)$ where $\tilde I_t$ is the partial summation of index  that will be used the following month
$$
\tilde I_{i} =    \sum_{j=1}^N a_j ( \tilde {S^j}^{l(k(i)+1),m(k(i)+1)}_{T(k(i)+1),i} - b_j) x^j_{i} 
$$
where
$$
\tilde {S^j}^{l,m}_{T,i} = \frac{1}{\tilde m} \sum_{q=T - \tilde l-i}^{\tilde m} S^j_{T - \tilde l-q}
$$
Using the last approximation we take into account the fact that a new index is under construction for the next period.
\end{itemize}
In the last case, the $I_t$ and $\tilde I_t$ are very correlated and the regression procedure using the Choleski method in the normal equation  (see \cite{bouchardwarin})
can fail when using many meshes in the last direction. That is the reason why we use only a linear approximation in the last direction.\\
\begin{remark}
The regression achieved is depending on the time step. If one regresses on $\hat Z_t= (S^0_t,I_t)$ for a current step after $t_{start}$,
the regression is only achieved on $S^0_t$ before  $t_{start}$ because $I_t$ does not exist. If one regresses on
$\hat Z_t= (S^0_t,I_t, \tilde I_t)$ for a current step between $t_{start}$ and $t_{end}$, $Z_t$ is restricted to $(S^0_t,\tilde I_t)$ before $t_{start}$
and $(S^0_t,I_t)$ after $T_{\hat n}$.
\end{remark}
Results obtained in optimization and simulation are given in table \ref{RegressTest}. Standard deviation of the portfolio with and without hedge is given.
In this test, we hedge each component of the index and the exchange rate.
\begin{table}[h]
\centerline{
%\begin{tabular}{|c|c|c|c|c|}  \hline
\begin{tabular}{|p{2cm}|p{2cm}|p{2cm}|p{2cm}|p{2cm}|}  \hline
%\noalign{\smallskip}\svhline\noalign{\smallskip}
  & Optimization  & Simulation &  \multicolumn{2}{|c|}{Standard deviation}  \\
& value &  value &  without hedge & with hedge \\  \hline
$\hat Z_t = S^0_t$  &  86.2 & 86.0   &  219.3   &   32.4       \\  \hline
$\hat Z_t = (S^0_t,I_t)$ & 97.7 & 97.4   & 204.1    &   21.3       \\  \hline
$\hat Z_t = (S^0_t,I_t, \tilde I_t)$ & 99.6  &  99.3 & 202.9  & 17.0  \\ 
\noalign{\smallskip}\hline\noalign{\smallskip}
 \end{tabular}}
\caption{Tests on regressor}
\label{RegressTest}
\end{table}
As scheduled, the regression on $ \hat Z_t= S^0_t$ give bad results in term of value compared to the other regressors and in term of hedge efficiency.
The regression on $\hat Z_t= (S^0_t,I_t)$ and  $\hat Z_t= (S^0_t,I_t,\tilde I_t)$ have a far higher value indicating a better optimization than in the first case.
The daily hedge is very efficient  dividing the standard deviation by $10$ for regressor 2 and even by $14$ for regressor 3.
\subsection{Components to hedge}
\begin{figure}[h]
\centering
\subfigure[ Hedging Brent position october 2006]
{\includegraphics[width=5.7cm]{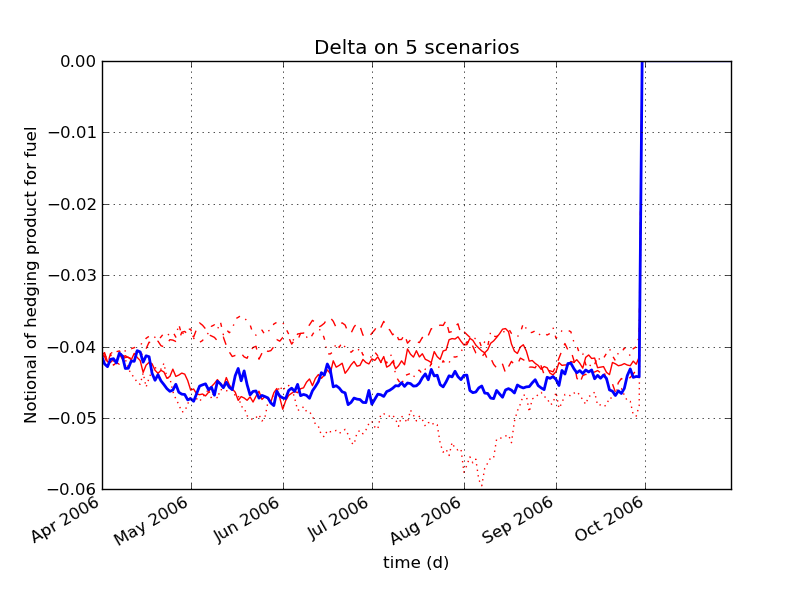}}
\subfigure[ Hedging Brent position january 2007]
{\includegraphics[width=5.7cm]{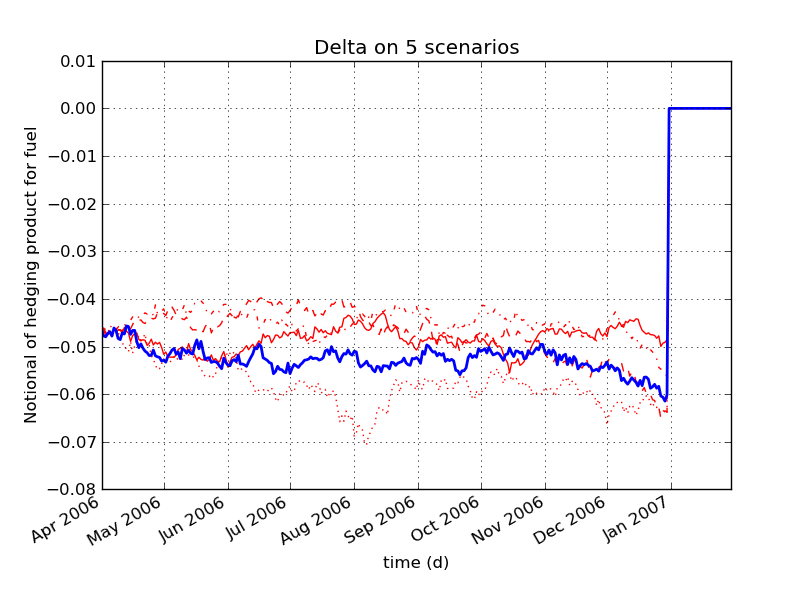}}
\subfigure[ Hedging Gas position january 2007]
{\includegraphics[width=5.7cm]{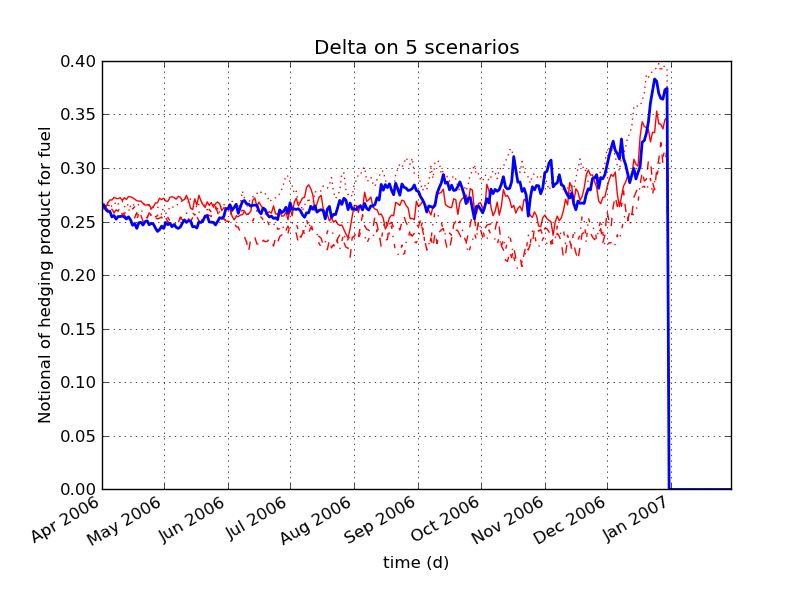}}
\subfigure[ Hedging exchange rate  position]
{\includegraphics[width=5.7cm]{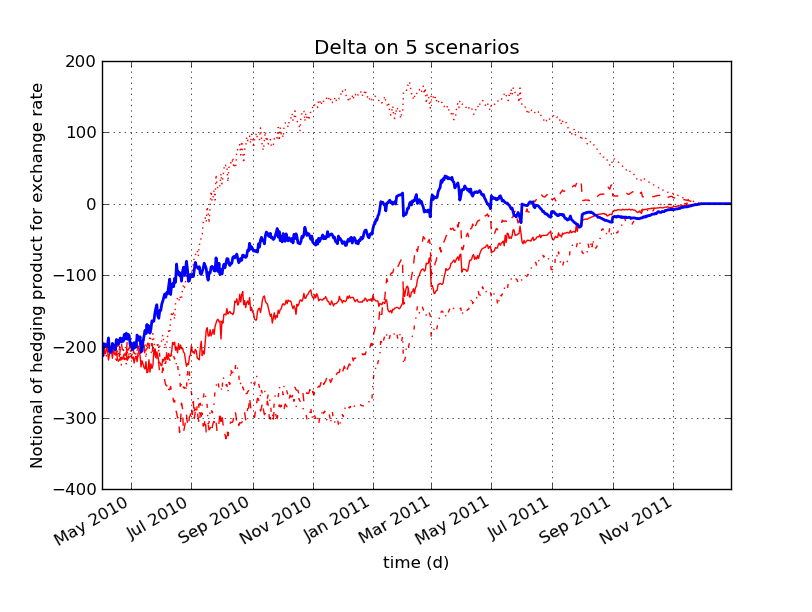}}
\caption{Hedging position for different maturities}
\label{figComp1}
\end{figure}

\begin{figure}[ht]
\centering
\includegraphics[width=12cm]{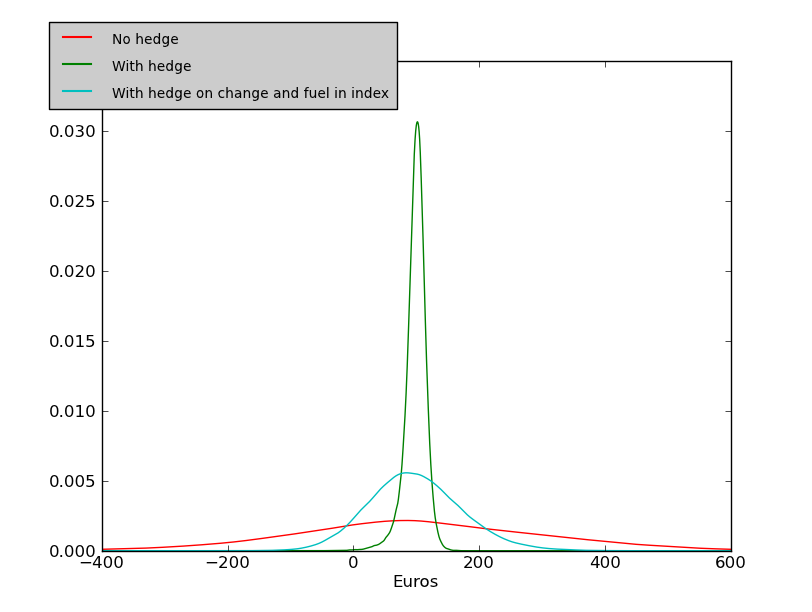}
%
%\vspace*{-0.25cm}
%
\caption{\label{Distribfig}  Test on the effectiveness of different hedges}
\end{figure}

In this subsection we choose to use a daily hedge calculated  with regressor 3.
On figure \ref{figComp1}, we give for the 5 same simulations :
\begin{itemize}
\item the hedging position in future october 2006 for Brent (so before the exercice have began) ,
\item the hedging position in Brent future for january 2007 for the same simulations,
\item the hedging  position for january 2007 in gas future,
\item the hedging position in foreign currency to hedge the product.
\end{itemize}
In table \ref{HedgeTest} we give the standard deviation obtained in simulation taking different hedge :
\begin{itemize}
\item for ``Spot gas'' we only hedge $S^0_t$ without taking into account the variation of the index components and exchange rate,
\item for ``Index components without exchange rate'' we only hedge against the variation of the values of the commodities without hedging against gas
spot variation and exchange rates variations,
\item for ``Index components and exchange rate'', we add to the previous hedge an hedge againt the exchange rate variations,
\item for ``Exchange rate'', we only hedge against the exchange rate variations,
\item for ``Total hedge'' we hedge against all the commodities and exchange rate in the contract.
\end{itemize}
On this example, it shows that the hedge has to be done against the variations  of the index components and that the hedge on gas spot variation is insufficient.
It also shows that all the commodities and exchange rate has to be hedged to get an efficient dynamic hedging.
\begin{table}[h]
\centerline{
\begin{tabular}{|p{5cm}|p{4cm}|}  \hline
 Hedge on &   Porfolio Standard deviation  \\ \hline
 Spot gas & 196.61 \\  \hline
Index component without exchange rate  &  77.20    \\  \hline
Index component and exchange rate   & 74.03    \\  \hline
Exchange rate & 201.7  \\  \hline
Total hedge  & 17.01 \\ 
\noalign{\smallskip}\hline\noalign{\smallskip}
 \end{tabular}}
\caption{Tests on hedge}
\label{HedgeTest}
\end{table}
On figure \ref{Distribfig}, we give de cash flow distribution obtained while not hedging, while hedging the index components, while hedging against all the product variations.

\subsection{Frequency of the hedge}
It is well know that the hedging error in Black Scholes framework converges to zero at a rate proportional to the square root of hedging frequency \cite{zhang, hayashi}.
Because the index is evolving quite smoothly, we wonder if we could hedge the derivative less frequently as for the index components : it could help us to decrease the computation time during optimization and would be interesting for practitioners because it would decrease the hedging cost due to illiquidity of the markets.
In table \ref{HedgeFreq}, we give the standard deviation of the portfolio hedge supposing we hedge all the component  but hedging the index with
different frequencies. It shows that the effectiveness of the hedge  decreases  very quickly with the hedge frequency.
\begin{table}[h]
\centerline{
\begin{tabular}{|p{4cm}|p{1.6cm}|p{1.6cm}|p{2cm}|p{2cm}|}  \hline
Frequency of the index hedge &   Every day & twice a week &  every week & twice a month \\ \hline
Standard deviation  & 17.0 &   42.9      &  54.78    &  57.5  \\  \hline
 \end{tabular}}
\caption{Standard deviation of the hedged portfolio with different hedge frequencies for the index components}
\label{HedgeFreq}
\end{table}
In table \ref{HedgeFreq7ap}, we give the standard deviation of the portfolio hedge supposing we hedge all the component  but hedging the index with
different frequencies before the first exercice date and balancing the hedging position of the index component every day after the first delivery day.
 \begin{table}[h]
\centerline{
\begin{tabular}{|p{4cm}|p{1.6cm}|p{1.6cm}|p{1.6cm}|p{2cm}|}  \hline
Frequency of the index hedge before $t_{start}$ & every week  &twice a month & every month & every quarter\\ \hline
Standard deviation  & 18.6  & 20.04  & 20.79  &  34.89 \\  \hline
 \end{tabular}}
\caption{Standard deviation of the hedged portfolio with different hedge frequencies for the index components before the first exercice date and hedging the index components
once a day after.}
\label{HedgeFreq7ap}
\end{table}
Results in table \ref{HedgeFreq7ap} clearly shows that the hedge frequency can be lowered before the first delivery date without affecting too much 
the hedging efficiency.

\section{Conclusion}
\label{secVI}
We have derived an efficient methodology to hedge dynamically some Swing product on gas market. We have shown that the strategy calculated
is very efficient for reducing the standard deviation of the portfolio for a realistic contract. 
Besides we have shown that on this contract the Swing value is mostly depending on the index component and that an efficient hedge has to deal with all
 the commodities involved in the index. At last, due to transaction cost it is always interesting to lower the hedge frequency and the last part of the study shows that is could be achieved before the first delivery date without losing too much the hedge efficiency.

\end{document}